\documentclass{elsart}
\usepackage{epsfig,amstext,amssymb}
\begin{document}

\begin{frontmatter}
\title{
The yield of air fluorescence induced by electrons
}

\author{F. Arqueros}\renewcommand{\thefootnote}{\fnsymbol{footnote}}\footnote{Corresponding author. \em{e-mail-address:} arqueros@gae.ucm.es},
\author{F. Blanco},
\author{A. Castellanos},
\author{M. Ortiz},
\author{J. Rosado}

\address{
Departamento de F\'{\i}sica At\'omica,
Molecular y Nuclear, Facultad de Ciencias F\'{\i}sicas, Universidad
Complutense, E-28040 Madrid, Spain.
}

\begin{abstract}
The fluorescence yield for dry air and pure nitrogen excited by
electrons is calculated using a combination of well-established
molecular properties and experimental data of the involved cross
sections. Particular attention has been paid to the role of
secondary electrons from ionization processes. At high pressure and
high energy, observed fluorescence turns out to be proportional to
the ionization cross section which follows the Born-Bethe law.
Predictions on fluorescence yields in
a very wide interval of electron energies (eV - GeV) and pressures
(1 and 1013 hPa) as expected from laboratory measurements are
presented. Experimental results at energies over 1 MeV are in very
good agreement with our calculations for pure nitrogen while
discrepancies of about 20\% are found for dry air, very likely
associated to uncertainties in the available data on quenching cross
sections. The relationship between fluorescence emission, stopping
power and deposited energy is discussed.
\newline
\end{abstract}
\end{frontmatter}

\section{Introduction}
\label{intro}
The fluorescence radiation from air showers generated by ultra-high
energy cosmic rays UHECRs provides a precise determination of the
longitudinal profile allowing the reconstruction of the primary
properties. This technique \cite{bunner} was first successfully used
by the Fly's Eye experiment \cite{flys_eye} and later by HiRes
\cite{HiRes1}. Fluorescence telescopes are being used by the Pierre
Auger Observatory \cite{auger1} to operate  simultaneously with a
giant air shower array. The planned Telescope Array project
\cite{TA} also relies in this technique. On the other hand, the
satellite-based experiments EUSO \cite{euso} and OWL \cite{owl} are
being designed for the detection of the fluorescence traces of air
showers viewing downward from the top of the atmosphere.
\par
The HiRes collaboration has reported measurements of the energy
spectrum of UHECRs \cite{HiRes2} which are in disagreement with
those of the AGASA air shower array \cite{agasa}. In order to
achieve a reliable energy calibration of a fluorescence telescope,
accurate values of the air fluorescence yield are required. This
need has promoted a number of experiments for the measurement of
this parameter in the wavelength interval of interest in this
technique, i.e. 300 - 400 nm. As is well known, in this spectral
interval, the fluorescence light comes from the First Negative
System of N$_2^+$ (B$^2\Sigma^+_u$ $\rightarrow$ X$^2\Sigma^+_g$)
and the Second Positive System of N$_2$ (C$^3\Pi_u$ $\rightarrow$
B$^3\Pi_g$) which in the following will be called 1N and 2P systems
respectively.
\par
Kakimoto et al. \cite{kakimoto} and Nagano et al.
\cite{nagano1,nagano2} have published measurements of the
fluorescence yield using electrons from the $\beta$ decay of a
$^{90}Sr$ radioactive source (average energy of about 0.85 MeV).
Experiments at higher energy are performed using linear accelerators
\cite{kakimoto}, \cite{ueno,air-fly,flash}. Finally, measurements at
low energy are being carried out with home made electron guns
\cite{ICRC2-UCM}. More information on the status of all these
experiments can be found in \cite{workshops}. On the other hand, the
effect of uncertainties in the fluorescence yield on the shower
reconstruction are being studied \cite{bianca,vitor}.

\par

Electrons passing through the atmosphere deposit energy due to
inelastic collisions with air molecules. Only a very small fraction
of these processes (i.e. excitation of B$^2\Sigma^+_u$ state of
N$_2^+$ and C$^3\Pi_u$ of N$_2$) gives rise to the production of
fluorescence light in our spectral range. However it is commonly
assumed that the fluorescence intensity is proportional to deposited
energy.
This assumption has not yet been established neither theoretically
nor experimentally, at least in a large energy range.

\par

In this paper, the air fluorescence yield is calculated using well
known theoretical properties and experimental data on the various
cross sections involved in the fluorescence emission. Particular
attention has been paid to the role of low energy secondary
electrons ejected in molecular ionizations.
As a result of our analysis recipes are given which allows the
calculation of the absolute value of the total fluorescence yield
(pressure and temperature dependent) as a function of electron
energy in a wide interval ranging from the threshold to the Gev
region. Comparison with available experimental data will be shown.

\section{Theoretical considerations}
\label{FY}
In this section well-established properties on the molecular
excitation (by electron collision) and de-excitation are reminded,
in particular, the relationship of the fluorescence yield with the
excitation cross section and the transition probabilities (radiative
and collisional). The contribution of secondary electrons to the
fluorescence emission is evaluated. The relationship between
deposited energy and fluorescence yield is discussed. All these
properties will allow us to make theoretical predictions on the
energy dependence of the fluorescence yield.

\subsection{Fluorescence yield and optical cross section}

Upon passage of an energetic electron through a molecular gas, the
{\sl fluorescence yield} for a band $\varepsilon_{vv'}$ is defined
as the number of fluorescence photons emitted in the molecular
transition $v-v'$ per incident electron and unit path length, where
$v$ and $v'$ represent the vibrational quantum numbers of the upper
and lower electronic states of the transition respectively. At very
low pressures and ignoring secondary processes, $\varepsilon_{vv'}$
is given by

\begin{equation}
\label{PY} \varepsilon_{vv'} = N\sigma_vB^{vv'} = N\sigma_{vv'}\,,
\end{equation}

where $N$ is the number of molecules per unit volume, $\sigma_v$ is
the excitation cross section for the $v$ level, $B^{vv'}$ is the
{\sl branching ratio} (i.e. the ratio between the partial
$A_r^{vv'}$ and total $A_r^v$ radiative transition probabilities of
the upper level)

\begin{equation}
\label{Bvv} B^{vv'} = \frac{A_r^{vv'}}{\sum{A_r^{vv'}}} =
\frac{A_r^{vv'}}{A_r^v}\,,
\end{equation}

and $\sigma_{vv'}$ is the so-called {\sl optical cross section}. As
is well known \cite{born-bethe}, the excitation cross sections
$\sigma_v$ are also approximately proportional to the Franck-Condon
factors $q_{X\rightarrow v}$, defined as the overlapping integrals
between the vibrational wavefunctions of the lower and upper levels
of the transition which in this case are the $X$ ground state of
N$_2$ and the $v$ excited state of N$_2$ or N$_2^+$ (depending on
the band system) respectively.

\par

Therefore

\begin{equation}
\label{FC} \frac{\sigma_{vv'}}{\sigma_{00}} =  \frac{q_{X\rightarrow
v}}{q_{X\rightarrow 0}}\frac{B^{vv'}}{B^{00}}\,.
\end{equation}

The above relation allows the calculation of any optical cross
section from data on a reference transition (e.g. $\sigma_{00}$). In
particular, for the nitrogen molecule the parameters
$q_{X\rightarrow v}$ and $B^{vv'}$ are available in the literature
\cite{gilmore,laux}.

\par

\subsection{Quenching, temperature dependence and pollutants effect}

At high pressure, molecular de-excitation by collision with other
molecules of the medium play an important role (collisional
quenching). The corresponding transition probability $A_c^v$ is
proportional to the collision frequency and thus to the gas pressure
$P$ (assuming fixed temperature). Defining $P'_v$ as the pressure
for which the probability of collisional quenching equals that of
radiative de-excitation $A_c^v(P'_v)=A_r^v$,

\begin{equation}
A_c^v (P) = A_r^v \frac{P}{P'_v}\,,
\end{equation}

so the fluorescence yield can be expressed in this case by

\begin{equation}
\label{PY_C} \varepsilon _{vv'}(P) = N\sigma_v
\frac{A_r^{vv'}}{A_r^v + A_c^v(P)} =
    N\sigma_{vv'} \frac{1}{1+P/P'_v}\,.
\end{equation}

\par

The characteristic pressure is given by

\begin{equation}
\label{p_par} P'_v =  \frac{\sqrt{\pi MkT}}{4\sigma_{nn}}
\frac{1}{\tau_r}\,,
\end{equation}

where $M$ is the molecular mass, $k$ is the Boltzmann's constant,
$T$ is the gas temperature and $\sigma_{nn}$ is the cross section
for collisional de-excitation.

\par

Collisional quenching enlarges the total transition probability $A^v
= A_r^v + A_c^v$ and therefore the lifetime of the population of
excited molecules $\tau^v = 1/A^v$ is shortened as compared with the
radiative one $\tau_r^v=1/A_r^v$ as

\begin{equation}
\frac{1}{\tau^v}=\frac{1}{\tau_r^v}+\frac{1}{\tau_c^v}\,,
\end{equation}

with $\tau_c^v=1/A_c^v$. Therefore lifetime depends on pressure as

\begin{equation}
\frac{1}{\tau^v(P)}=\frac{1}{\tau^v_r}(1+\frac{P}{P'_v})\,.
\end{equation}

\par

Both $\tau_r^v$ and $P'_v$ can be measured in a plot of reciprocal
lifetime versus pressure (Stern-Volmer plot). This is a very well
established technique in use since many years ago for the
experimental determination of radiative lifetimes and quenching
cross sections.

\par

For gas mixtures like air, the above formulation is valid taking
into account that in equation (\ref{PY}) and (\ref{PY_C}) $N$ is the
density of fluorescence emitters (e.g. nitrogen amounts a 79\% of
the molecular density in air) and the quenching of excited molecules
is due to collisions with the various components. Neglecting other
components apart from nitrogen and oxygen the $P'_v$ parameter for
air is written as

\begin{equation}
\label{p_par2} P'_v =  \frac{\sqrt{\pi M_nkT}}{4\tau_{r}}
\{f_n\sigma_{nn} + f_o\sigma_{no} \sqrt{\frac{M_n+M_o}{2M_o}}
\}^{-1}\,,
\end{equation}

where $M_n$ and $M_o$ are the molecular masses of nitrogen and
oxygen respectively, $\sigma_{nn}$ and $\sigma_{no}$ are the cross
sections for collisional quenching of the excited nitrogen with
nitrogen and oxygen molecules respectively. The parameters $f_n$ and
$f_o$ are the molecular number fraction of nitrogen and oxygen in
air.

\par

In some media (e.g. plastic scintillators), excited molecules may
decay by other additional processes like intersystem crossing or
internal conversion \cite{birks} and thus the lifetime is further
shortened. However for nitrogen (air) gas such internal processes do
not take place and therefore they will be neglected in this
analysis.

\par

Equations (\ref{p_par}) and (\ref{p_par2}) give the dependence of
fluorescence yield with temperature. Firstly, $P'_v$ shows a $\sqrt
T$ dependence (i.e. collision frequency). On the other hand, the
quenching cross sections $\sigma_{nn}$ and $\sigma_{no}$ are
expected to depend on the kinetic energy of the colliders and
therefore a further contribution to the temperature dependence could
be expected. Unfortunately no data on this dependence for air in our
temperature range is available in the literature and thus this
effect, which is not expected to be relevant, will be neglected in
our analysis. In fact measurements from the AIRFLY experiment are
consistent with quenching cross sections independent on temperature
\cite{air-fly}.

\par

As is well known oxygen is responsible for the large quenching of
air fluorescence as compared with the smaller auto-quenching of pure
nitrogen ($\sigma_{no}>> \sigma_{nn}$). Contribution of other air
components (e.g. water vapor) or pollutants can be easily evaluated
with equations (\ref{PY_C}) and the appropriate extension of
(\ref{p_par2}) as far as the corresponding quenching cross sections
are available. In this paper quenching by other air components will
not be studied.

\subsection{Contribution of secondary excitations}

Apart from direct excitation by electron collision, the upper level
can be populated by de-excitation of higher energy electronic
states. This cascading effect contributes to the fluorescence yield
enlarging the apparent optical cross sections, so experiments
devoted to determine excitation cross sections need to make the
appropriate corrections to the measured apparent values. While for
our problem, this apparent cross section is the parameter of
interest, no significant cascade effects have been reported and so
this will be neglected in this analysis.

\par

Pure vibrational transitions (i.e. inside a given electronic state)
induced by collisions with other molecules of the medium can
redistribute somewhat the population of the upper levels. Very
scarce data on this process are available in the literature. However
the effect on the total fluorescence yield $\sum_{vv'}{\varepsilon
_{vv'}}$ (i.e. the total number of photons emitted per electron per
meter) is expected to be negligible with a possible small effect on
the spectrum of the fluorescence radiation. This effect will not be
taken into account in our analysis.

\par

A very important secondary contribution will be that from low energy
electrons ejected in ionizations. These excite molecular nitrogen
increasing the observed fluorescence \cite{blanco}. As
aforementioned, for each primary incident electron of kinetic energy
$E$, the number of direct excitation processes to the $v$ level per
unit length is $N\sigma_v(E)$. On the other hand, the number of
generated secondary electrons per unit length is $N\sigma_{ion}(E)$
where $\sigma_{ion}(E)$ is the ionization cross section. For our
calculations we will use the so-called gross ionization cross
section
which also includes processes in which more than one electron is
generated in a primary interaction (even though contribution from
those processes is small). If we name $\alpha_v(E,P)$ the fraction
of these secondary electrons that excite the upper v level
($\alpha_{vv'}=\alpha_v B^{vv'}$ the fraction that excites $v-v'$
band emission), then an effective optical cross section $\sigma
^{eff}_{vv'}$ can be defined as
\begin{equation}
\label{sigma-eff} \sigma ^{eff}_{vv'} = \sigma_{vv'} (E) +
\alpha_{vv'}(E,P)\sigma_{ion}(E)\,,
\end{equation}

and the total fluorescence yield given by equation (\ref{PY_C})
becomes

\begin{equation}
\label{PY_C-eff} \varepsilon _{vv'}(P) = N \frac{1}{1+P/P'_v}\:
\sigma_{vv'}^{eff}\,.
\end{equation}

The effect of these secondary electrons results the dominant
mechanism for the production of 2P fluorescence at high pressure
\cite{blanco}. In addition, as will be shown in sub-section 3.1, a
non-negligible contribution of the 1N fluorescence is due to these
secondary electrons.

\par

The $\alpha_{vv'}$ parameter can be calculated by means of a Monte
Carlo simulation. For this work the simple algorithm described in
\cite{blanco} has been used with a somewhat improved treatment of
the electrons escaping the interaction region. The resulting small
dependence of $\alpha_{vv'}$ on primary electron energy, pressure
and characteristic size of interaction region $R$, can be
approximated (for 1hPas and higher pressures) by

\begin{equation}
\label{alpha_00} \alpha _{vv'}(E,P) = \min \{s_0\ln\frac{P\times
R}{s_1}, e_0\ln\frac{E}{e_1}\}\,.
\end{equation}

This expression is purely empirical (a fit to the MC results). The
value of the  $s_0$, $s_1$, $e_0$ and $e_1$ parameters depend on the
particular case. Results for 0-0 bands of 1N and 2P systems are
shown in Table 1.

\par

The two terms inside the {\sl min} function in equation
(\ref{alpha_00}) can be easily interpreted as corresponding
respectively to those conditions where electrons can escape from the
interaction region (low $P\times R$ values) and those where
electrons lose all their energy inside the interaction region (high
$P\times R$ values). While in the first case fluorescence by
secondary electrons is limited by the traversed material, in the
second one no (primary or secondary) electron escapes from the
interaction region and fluorescence contribution is given by the
primary incident energy. In both cases the (very slow) logarithmic
dependence arises from the very small dependence of the
secondary-electrons energy-distribution on the primary electron
energy. For a given primary energy, the saturation $PR$ value (for
which conditions change from one to the other behavior) can be
interpreted as the range of secondary electrons in the gas and, as
can be easily checked, this is almost independent of the studied
process (1N or 2P fluorescence). A paper describing more details of
this model is in preparation \cite{blanco2}.

\par

Since $\alpha _{vv'}$ is proportional to the corresponding optical
cross section, expression (\ref{FC}) leads to

\begin{equation}
\label{alpha} \frac{\alpha_{vv'}}{\alpha_{00}} =
\frac{q_{X\rightarrow v}}{q_{X\rightarrow
0}}\frac{B^{vv'}}{B^{00}}\,,
\end{equation}

which allows the calculation of the various $\alpha_{vv'}(E,P)$
parameters from  $\alpha _{00}$.

\begin{center}
\begin{tabular}{c||c|c|c|c|}

    &  $s_0$   & $s_1$ (hPa$\times$cm)   & $e_0$   & $e_1$(eV)   \\
\hline \hline
1N (0-0) & $8.67 \cdot 10^{-3}$ & $1.92 \cdot 10^{-2}$  & $1.37 \cdot 10^{-2}$ & 73.7    \\
\hline
2P (0-0) & $2.00 \cdot 10^{-3}$ & $1.36 \cdot 10^{-5}$ & $3.22 \cdot 10^{-3}$ &  0.942   \\
\hline
\end{tabular}
\end{center}

\small
Table 1.-  {\sl Values of the parameters in equation
(\ref{alpha_00}) for the most prominent bands of the 1N and 2P
molecular systems.}
\normalsize

\par\vspace{0.5cm}

Note that experiments determining quenching cross sections from the
pressure dependence of fluorescence intensity may overestimate the
$P'$ values if they ignore the effect of secondary electrons by
using equation (\ref{PY_C}) instead of the (\ref{PY_C-eff}) right
one. As mentioned above the $P'$ parameter can be also measured in a
plot of reciprocal lifetime versus pressure. This is a safe
technique not affected by secondary electrons.

\subsection{Stopping cross section and deposited energy}

The stopping power for electrons $S$, i.e. the energy loss per unit
length of traversed matter due to both excitation and ionization
processes, can be accurately determined for energies above 1 keV by
the well known Bethe-Bloch formula \cite{bethe-bloch1}.  At lower
energy, experimental results are available down to 25 eV
\cite{roldan}. On the other hand, in general, $S$ can be expressed
\cite{born-bethe} as

\begin{equation}
\label{sigma_inel} S = \frac{dw}{dx} = N\sum E_{n}\sigma_{n}\,,
\end{equation}

where the summation includes the energy $E_n$ and cross section
$\sigma_n$ for all kinematically accessible excited states n
(ionization processes contributing there with a large family of
continuum states). The cross section $\sigma_n$ depends on the
electron incident energy, and can be determined in electron
scattering experiments \cite{williart}. A {\sl stopping cross
section} $\sigma_{st}(E)$, proportional to $S$, can be easily
defined \cite{born-bethe} by rewriting (\ref{sigma_inel}) as

\begin{equation}
\label{sigma_inel_2} S = \frac{dw}{dx} = NR_y\sum
(E_{n}/R_y)\sigma_{n} = NR_y\sigma_{st}\,,
\end{equation}

where $R_y$ is the Rydberg constant (13.606 eV). This
$\sigma_{st}(E)$ cross section, proportional to the stopping power,
will be useful for comparison purposes in the present analysis.

\par

The {\sl fluorescence efficiency} $\Phi_{vv'}$
\cite{nagano1,nagano2}, \cite{sakaki} is defined as the fraction of
deposited energy which is emitted as photons of the $v-v'$ band.
Slow electrons deposit all the energy locally and thus the
Bethe-Bloch formula gives the dependence of deposited energy on $E$.
Therefore for low energy electrons

\begin{equation}
\label{phi} \Phi_{vv'}(P)=\frac{E_{vv'}\varepsilon _{vv'}}{dw/dx}
=\frac{E_{vv'}\sigma_{vv'}}{R_y\sigma_{st}}\frac{1}{1+P/P'_v}\,.
\end{equation}

\begin{figure}[h]
\centering \epsfig{file=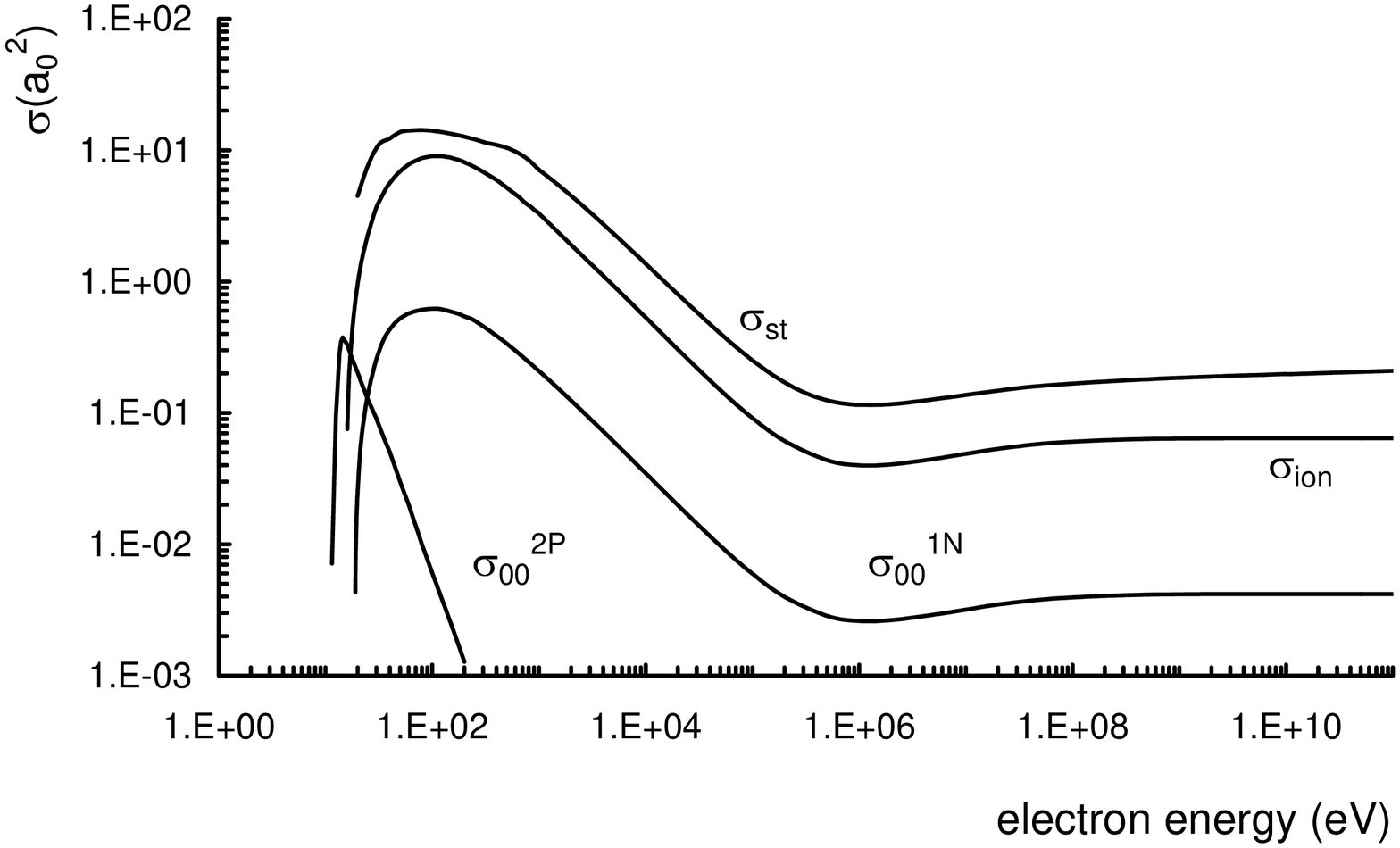,width=1.0\linewidth} \vskip
-1.2cm \caption{{\sl Cross sections (atomic units) involved in the
emission of fluorescence as a function of energy for air at
atmospheric pressure. The optical cross section $\sigma_{00}^{1N}$
for the 1N system has been obtained from experimental data at low
energy assuming a Born-Bethe behavior at high energy and including
the density correction (see 3.1). The ionization cross section
$\sigma_{ion}$ has been taken from \cite{s_ion}. Experimental
results have been used for the 2P optical cross section
$\sigma_{00}^{2P}$ (see 3.2). The $\sigma_{st}$ cross section
(proportional to the electron stopping power) has been obtained from
\cite{bethe-bloch1} and \cite{roldan} for energies above and below 1
keV respectively. Note that $\sigma_{st}$ and $\sigma_{ion}$ follow
a similar behavior although they are not exactly proportional.}}
\end{figure}

Note that the interpretation of the fluorescence efficiency given by
some authors \cite{bunner,nagano1} is different from that of
equation (\ref{phi}). These authors assume that all the electron
energy loss is employed in the excitation of the molecules to one
single upper level, a fraction is supposed to be emitted as
fluorescence radiation and the remaining energy is lost by internal
quenching. Such a model which is valid in some media (e.g. plastic
scintillators) \cite{birks} is not applicable in our case. In fact,
only a small fraction of the lost energy is used in the excitation
of the level $v$ of interest while the remaining energy gives rise
to other direct excitations, not internal processes. The authors of
ref. \cite{nagano1} argued recently along similar lines
\cite{sakaki}.

\par

At high energy the stopping power also accounts for the production
of high energy secondary electrons which do not deposit their energy
locally (large range and bremsstrahlung emission). Unless the
interaction region (e.g. gas cell in a lab experiment) is very
large, a non-negligible fraction of the energy transferred to the
medium by the primary electron is not deposited inside the volume
and consequently a non-negligible fraction of the fluorescence
radiation is not emitted in the interaction region.

\par

According to equation (\ref{PY_C-eff}), several terms contribute to
the total fluorescence yield. At low energy and low pressure the
main contribution is direct excitation. In principle the energy
dependence of $\sigma_{st}$ and $\sigma_{vv'}$ are not expected to
be the same since $\sigma_v$ excitation cross section only accounts
for one of the many inelastic processes involved in the energy loss.
On the other hand, even if $\sigma_n$ followed the same behavior for
all processes (including those leading to the generation of
fluorescence light) $\varepsilon_{vv'}$ is not expected to be
proportional to $\sigma_{st}$ because of the different weighting
values of $E_n$ in equation (\ref{sigma_inel_2}) depending on each
particular process. In Figure 1 the cross sections involved in
fluorescence emission and energy deposition are shown as a function
of energy for comparison. For instance, the different behavior of
$\sigma_{st}$ and 2P optical cross section is evident.

\begin{figure}[h]
\centering \epsfig{file=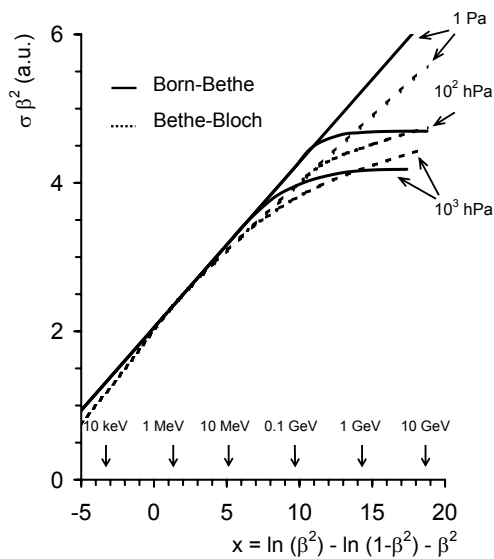,width=0.6\linewidth} \vskip
-1.5cm \caption{{\sl Comparison on a Fano plot of the stopping cross
section (Bethe-Bloch) and the ionization cross section (Born-Bethe)
for air at 1Pa, 100 hPa and atmospheric pressure. Cross sections (in
arbitrary units) have been normalized in order to get a common value
at 1 MeV for all cases. Pure Born-Bethe function (e.g. ionization
cross section at 1Pa
) is linear while the Bethe-Bloch stopping cross section clearly
deviates from linearity. Density effect tends to saturate both
functions at high energy although the effect is more pronounced for
the ionization cross section.}}
\end{figure}

As shown in detail later, at high energy and high pressure the
fluorescence yield is approximately proportional to the ionization
cross section which follows the Born-Bethe relativistic law
\cite{born-bethe}. The Born-Bethe law for the electron-molecule
collision can be expressed as

\begin{equation}
\label{born-bethe} \sigma=\frac{A}{\beta^2} \{\ln C\beta^2-\ln
(1-\beta^2)-\beta^2 \}\,,
\end{equation}

where $\beta$ is the relativistic speed and, $A$ and $C$ are
constants. At very high energies, a density correction has to be
applied to take into account the polarization of the molecules along
the electron path \cite{salvat,seltzer,perkins}. For that purpose,
we have included the density correction in the transversal
contribution of the Born-Bethe expression \cite{born-bethe}, in the
form proposed by \cite{salvat}. The resulting expression turns out
to be

\begin{equation}
\label{born-bethe2} \sigma=\frac{A}{\beta^2} \{\ln C\beta^2-\ln
(1-\beta^2)-\beta^2 -\delta_F\}\,,
\end{equation}

where $\delta_F$ is the usual density correction parameter included
in the Bethe-Bloch formula for the stopping power \cite{leo}.
Because of this density correction, the ionization cross section of
the nitrogen molecule at high energy depends on pressure.
\par
For any pure Born-Bethe cross section (\ref{born-bethe}), $\sigma
\beta^2$ versus $x=\ln (\beta^2)-\ln (1-\beta^2)-\beta^2$  is a
straight line (the so-called Fano plot). In Figure 2 the ionization
cross section has been represented on a Fano plot for several
pressures together with the corresponding stopping cross section
(Bethe-Bloch formula) \cite{bethe-bloch2}. As expected, at very low
pressure, $\sigma_{ion}$ follows a linear behavior while
$\sigma_{st}$ clearly deviates from linearity. At high pressure and
very high energy both functions flattens as expected from the
density effect but following a slightly different behavior. On the
other hand Figure 2 shows that at low energy $\sigma_{st}$ is not
exactly proportional to $\sigma_{ion}$.

\par

In summary we can conclude that fluorescence yield is not expected
to be exactly proportional to deposited energy since most of the
fluorescence radiation is generated by low energy electrons (either
primaries or secondaries) for which fluorescence emission
($\sigma_{1N}$, $\sigma_{2P}$) and stopping power ($\sigma_{ion}$)
follow different energy behavior.

\section{The fluorescence yield in pure nitrogen and dry air}
\label{FY_N_A} In this section experimental data on the
excitation/de-excitation of the upper levels of the 1N and 2P
systems of nitrogen will be analyzed. This information will allow us
the evaluation of the fluorescence yield. For all the calculations
shown in this section a temperature of 300 K has been assumed. The
density of fluorescence emitters (nitrogen molecules) $N$ in dry air
has been assumed to be 79\% of the total molecular density.

\subsection{The First Negative system}
\label{FY_1N} The 1N system of N$_2^+$ has been studied by many
authors. In particular many measurements of the optical cross
section $\sigma_{00}$ corresponding to the strongest band at 391.2
nm are available in the literature. At very low energies, absolute
measurements have been reported by Borst and Zipf \cite{borst} (19
eV $<E<$ 3.0 keV), Stanton and St John\cite{stanton} (19 eV $<E<$
400 eV), Aarts et al. \cite{aarts} (97 eV $<E<$ 6 keV), Srivastava
and Mirza \cite{srivastava} (70 eV $<E<$ 4 keV), Holland
\cite{holland} (90 ev $<E<$ 2 keV), McConkey and Latimer
\cite{mcconkey1} (26 eV $<E<$ 300 eV) and McConkey et al.
\cite{mcconkey2} (18 eV $<E<$ 2 keV). In Figure 3 a significant set
of the above measurements have been represented on a Fano plot.
Also, relative measurements published in \cite{arqueros}, normalized
to an average value of the above authors in the overlapping region,
have been included.

\begin{figure}[htb]
\centering \epsfig{file=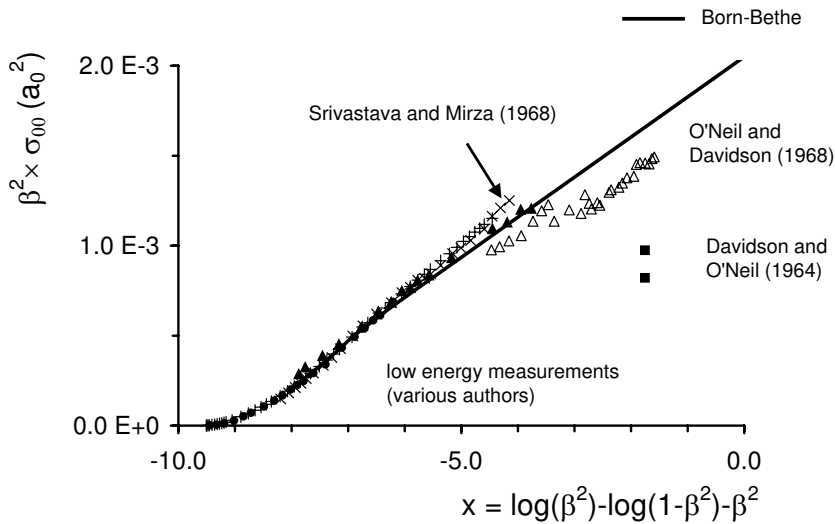,width=1.0\linewidth} \vskip
-1.0cm \caption{{\sl Fano plot for the optical cross section
$\sigma_{00}$ (in atomic units) of the 1N system. Experimental
results (symbols) show a linear behavior at energies over about 150
eV. Discrepancies in the slope are very likely due to calibration
factors. An average linear fit (continuous line) has been used in
this work for the calculation of the 1N fluorescence yield.}}
\end{figure}

At higher energies some measurements are available. Data of O'Neil
and Davidson in the range 3 - 60 keV \cite{davidson2} are shown in
the figure. Cross sections inferred from previous measurements of
these authors with 50 keV electrons at high pressure
\cite{davidson1} are significantly smaller and they will not be used
in the present analysis. As shown in the figure, the above
measurements follow a linear behavior in a Fano plot. Note that the
density correction is only relevant at very high energy. The slope
of the various data sets, i.e. $A$ parameter in equation
(\ref{born-bethe}), have slightly different values while a
reasonable agreement in the abscise in the origin is found, i.e. $C$
parameter. This features can be interpreted as discrepancies in the
absolute calibration factors of the various experiments, while their
results agree in the shape of the $\sigma_{00}(E)$ function.

\par

An average linear fit of the available data to equation
(\ref{born-bethe}) is shown in Figure 3. The values of the
parameters were found to be $A = 2.230 \times 10^{-4}$ a$_0^2$ and
$C = 9.826 \times 10^{3}$. None of these experimental works reports
any contribution of cascades to the measured optical cross section.
As a further test of the absence of cascades, it has been checked
\cite{blanco} that the measured optical cross sections of this
system fulfills equation (\ref{FC}). Therefore the above Born-Bethe
function provides us with a reliable value of $\sigma_{00}$ in the
whole energy range (see Figure 1). Note that at very high energy the
density correction has to be included. Optical cross section for
other 1N bands can be easily obtained using relation (\ref{FC}).

\par

While at very low pressures and low energies $\sigma_{vv'}$ values
suffice for evaluating fluorescence yields by means of equation
(\ref{PY}), at those conditions we are interested here, quenching
and secondary-electron contributions have to be considered by means
of equation (\ref{PY_C-eff}). Expression (\ref{alpha_00}) and Table
1 show that the $\alpha_{00}$ parameter, accounting for the
contribution of secondary electrons of the most intense band of the
1N system (391.2 nm), ranges from 0 (at low $E$ and/or low $P\times
R$ conditions) to 0.12 (at $E$ = 1MeV and $P\times R$=5
atm$\times$cm). In order to appraise the importance of these
secondary contributions it must be noted that $\sigma_{00}\simeq
0.075 \sigma_{ion}(E)$ for 1N system at large energies (see Figure
1). Introduction of that result in equation (\ref{PY_C-eff}) reveals
that secondary/direct contributions are in the $\alpha_{00}/0.075$
ratio and thus, contributions from secondary electrons can be very
important at high pressures and high energies.

\begin{figure}[htb]
\centering \epsfig{file=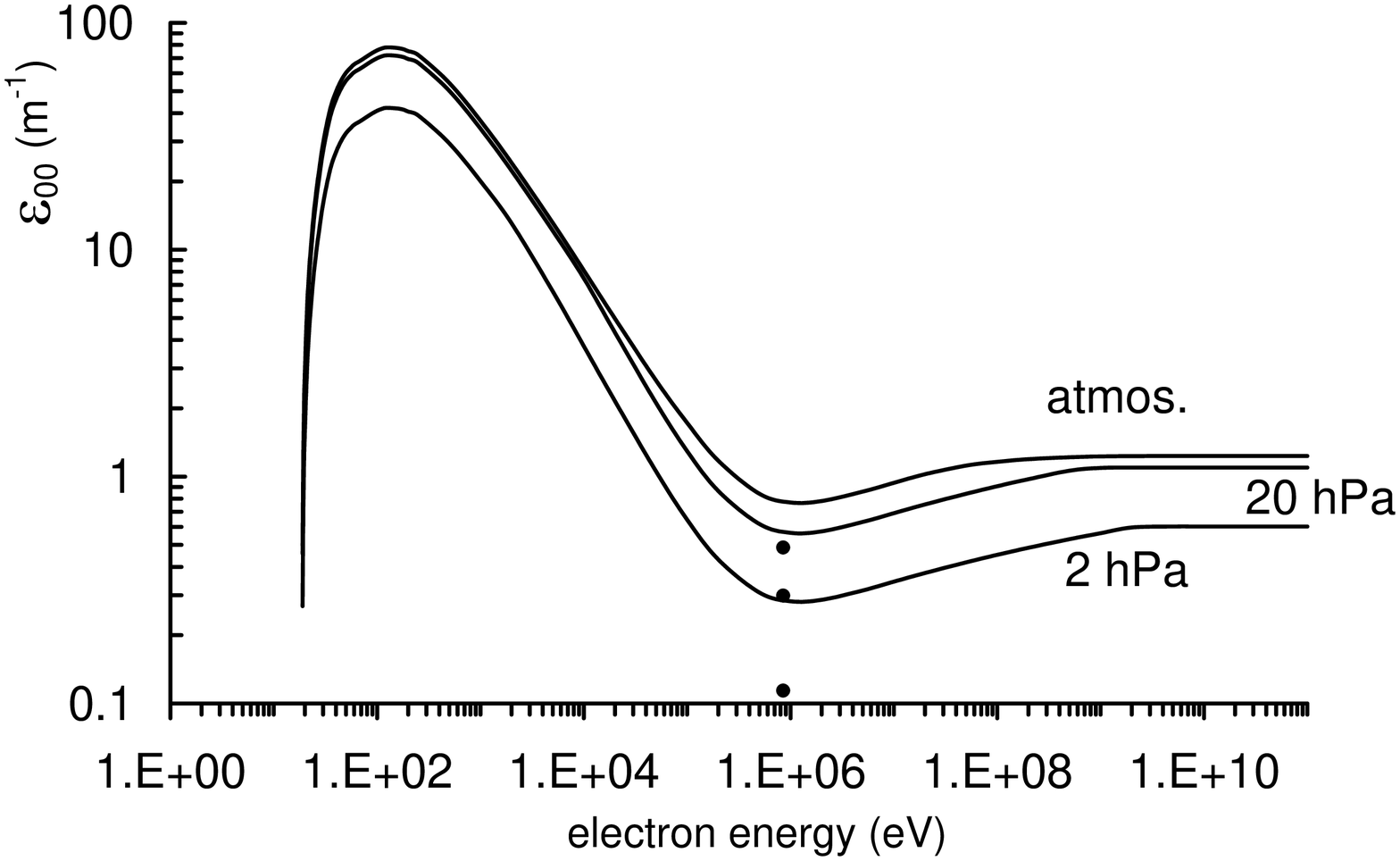,width=0.7\linewidth} \vskip
-0.7cm \caption{{\sl Fluorescence yield of the most prominent band
of the 1N System (391.2 nm) for pure nitrogen at several pressures.
Predictions of this work (continuous line) are compared with
experimental data of Nagano et al. \cite{nagano2} ($\bullet$) at the
same pressures.}}
\end{figure}

\begin{figure}[h]
\centering \epsfig{file=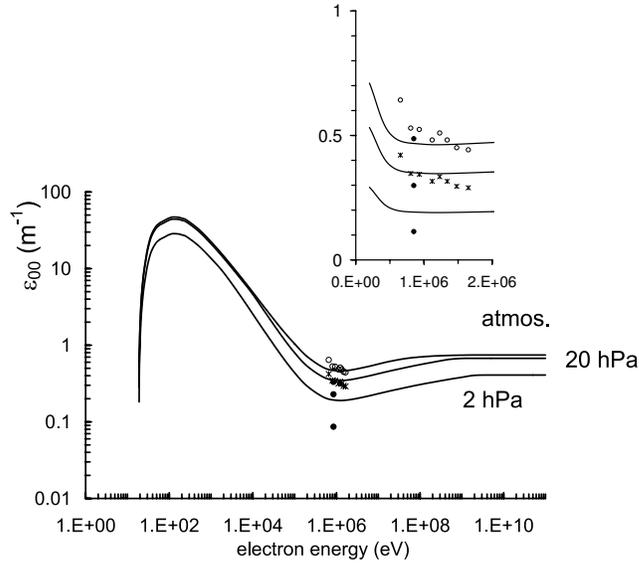,width=0.6\linewidth} \vskip
-0.1cm \caption{ {\sl Same as Figure 4 for dry air. In addition,
experimental values of Hirsh et al. \cite{hirs} at 20 hPa ($\ast$)
and atmospheric pressure ($\circ$) are shown for comparison (see the
inset for a more detailed comparison).}}
\end{figure}

The above results allow the calculation of $\varepsilon_{00}$ as a
function of electron energy, pressure, temperature and
characteristic size of the interaction region in a given
experimental arrangement. As an example Figures 4 and 5 show
$\varepsilon_{00}$(E) for pure nitrogen and dry air respectively, in
a wide energy range (eVs - GeVs) for several pressures (2, 20 and
1013 hPa) and a temperature of T = 300 K. Characteristics pressures
of $P'_0$ = 1.7 and 1.3 hPa for pure nitrogen and dry air
respectively have been assumed (see sub-section 3.3 for a discussion
on available $P'$ values). In these calculations a characteristic
size of $R$ = 2.5 cm has been assumed, which can be considered as
representative for typical lab measurements. The fluorescence yield
results of Hirsh et al.\cite{hirs}, for dry air, and Nagano et al.
\cite{nagano2}, for both nitrogen and dry air, are shown in the
figures for comparison. Direct measurements of \cite{hirs} are in
agreement with our predictions showing that our correction for
secondary electrons accounts for these experimental results. On the
contrary, our calculations are in clear disagreement with
measurements of \cite{nagano2} for which a two line analysis had to
be applied in order to subtract the contribution of other band not
separated by the spectroscopy filter. As shown below, the
contribution of the 1N system to the total air fluorescence yield
(at high pressure) is in the range of 15\% and therefore the above
discrepancy is not so relevant for the total fluorescence yield.


\subsection{The Second Positive system}
\label{FY_2P} Fluorescence of the 2P system in both pure nitrogen
and air has been extensively studied by many authors. Experiments
aiming at measuring the electron excitation cross section of the
upper level C$^3\Pi_u$ carried out on pure nitrogen at low electron
energy ($E <$ 1 keV) and low pressure (a few mTorr or lower) show a
fast $E^{-2}$ dependence of the fluorescence light
\cite{fons,shemansky,imami} much faster than that of the 1N system.
This result is expected, taking into account the optically-forbidden
nature of the 2P transition. In Figure 1 the optical cross section
for the $0-0$ band of the 2P system from the abovementioned
measurements is shown.

\begin{figure}[htb]
\centering \epsfig{file=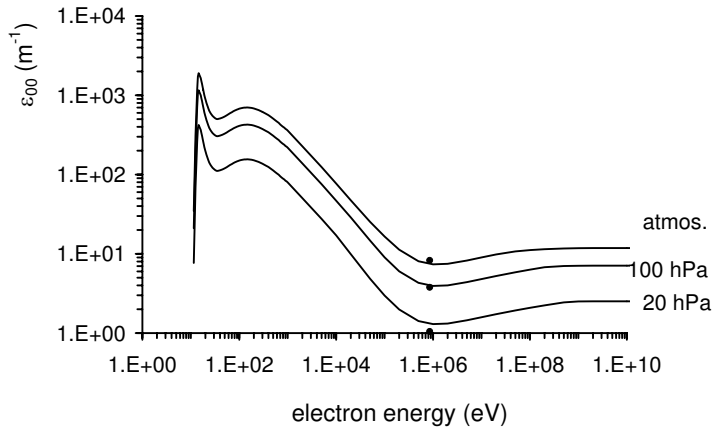,width=0.8\linewidth} \vskip
-0.5cm \caption{{\sl Fluorescence yield of the most prominent band
of the 2P System (337.0 nm) for pure nitrogen at several pressures.
Predictions of this work (continuous line) are compared with
experimental data of Nagano et al. \cite{nagano2}.} }
\end{figure}

Since at 1 keV energies, 2P fluorescence observed at low pressure is
much weaker than that of the 1N system, a naive extrapolation would
predict that the 2P fluorescence should be completely negligible
compared to 1N one at 30 keV energies and above. On the contrary
experiments \cite{kakimoto,nagano1,nagano2}, \cite{davidson1}
carried out at high energies (up to 1 GeV) on air at high pressure
(ranging from a few hPa up to 1 atm) shows that the 2P fluorescence
even dominates over the 1N system. This feature can be explained
taking into account the effect of the secondary electrons ejected in
ionization processes \cite{blanco}. The contribution of these
secondary electrons have been evaluated from a MC calculation as
explained above and the values of the corresponding parameters of
expression (\ref{alpha_00}) for the most intense band of this system
(337.0 nm) are shown in Table 1.

\begin{figure}[htb]
\centering \epsfig{file=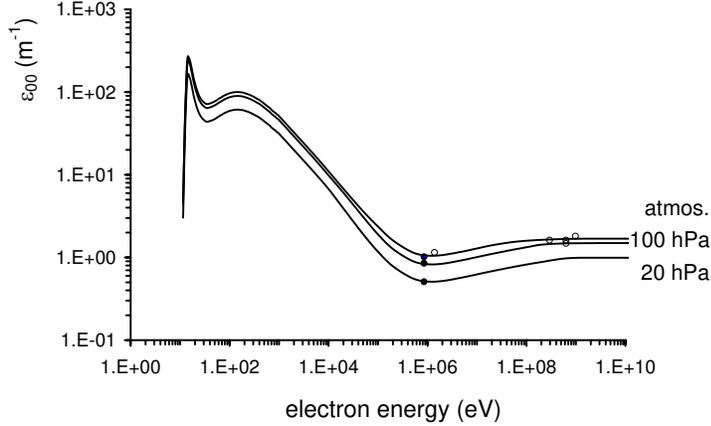,width=0.8\linewidth} \vskip
-0.5cm \caption{{\sl Same as Figure 6 for dry air. In addition
experimental results of Ueno \cite{ueno} at atmospheric pressure
($\circ$) are shown.}}
\end{figure}

Characteristic pressures of $P'_0$ = 77 and 13.1 hPa for pure
nitrogen and dry air respectively have been assumed (see below for a
discussion on available $P'_v$ values). Figures 6 and 7 show the
predicted fluorescence yield for the 0-0 band of the 2P system as a
function of electron energy from threshold (eVs) up to the GeV
region at several pressures in pure nitrogen and dry air
respectively. An interaction region of 2.5 cm radius has also been
assumed for this calculation. Experimental data for the 337.0 nm
band at 0.85 MeV \cite{nagano2}, 1 MeV and near 1 GeV \cite{ueno}
are shown for comparison.


\subsection{Total fluorescence yield}
\label{TFY}
The total fluorescence yield, i.e. the sum of the contributions from
all the molecular bands of both systems 1N and 2P,

\begin{equation}
\label{TFY1} \varepsilon={\sum_{vv'}{\varepsilon_{vv'}^{1N}}} +
{\sum_{vv'}{\varepsilon_{vv'}^{2P}}}
\end{equation}

can be easily calculated from (\ref{PY_C-eff}) provided that the
involved molecular parameters are available for all transitions.
Taking into account (\ref{FC}) and (\ref{alpha}) the contribution of
each band system to the total fluorescence yield can be expressed as
a function of a reference transition (e.g. 0-0) as

\begin{equation}
\label{TFY2} \varepsilon^{system} =
N\frac{\sigma^{eff}_{00}(E,P)}{q_{X\rightarrow 0}\,B^{00}}
{\sum_{vv'} q_{X\rightarrow v}B^{vv'}\frac{P'_v}{P + P'_v}}=
N\frac{\sigma^{eff}_{00}(E,P)}{q_{X\rightarrow 0}\,B^{00}} {\sum_{v}
q_{X\rightarrow v}\frac{P'_v}{P + P'_v}}\,,
\end{equation}

where (\ref{Bvv}) has been used. All the parameters depend on the
{\sl system}, 1N or 2P. The terms inside the v-v' sum represent the
relative intensities of the bands for the corresponding system and
at high pressures ($P\gg P'_v$) they can be approximated by

\begin{equation}
\label{rel_int} \varepsilon_{vv'} \sim q_{X\rightarrow v} B^{vv'}
P'_v\,.
\end{equation}

In general, equation (\ref{rel_int}) is able to predict experimental
results on relative intensities although a detailed comparison is
beyond the scope of this paper.

\small
\begin{center}
\begin{tabular}{c||c|c|c|c|c|c||c||c|c|}

\hline
v/\ v' & 0 & 1 & 2 & 3  & 5 & 5 & $q_{X\rightarrow v}$ & $P^{v}_{nitr}$ (hPa) & $P^{v'}_{air}$ (hPa)\\
\hline
 & 391.2 & 427.5 & 470.6 & 522.5 & 586.1 & 665.9 & & &    \\
0 \hspace{0.5cm}  & 0.627 & 0.204 & 0.043 & 0.007 & 0.001 & 0.000 & 0.883 & 1.7  & 1.3 \\
\hline
 & 358.0 & 388.2 & 423.4  & 464.9 & 514.6 & 575.0 & &         &               \\
1  \hspace{0.5cm} & 0.041 & 0.029 & 0.030  & 0.011 & 0.003 & 0.001 & 0.114 & --  & -- \\
\hline
 & 330.5 & 356.1 & 385.5  & 419.7 & 459.7 & 570.4 & &         &             \\
2  \hspace{0.5cm} & 0.000 & 0.001 & 0.000  & 0.000 & 0.000 & 0.000 & 0.002 & -- & --  \\
\hline

\end{tabular}
\end{center}

Table 2.- {\sl Molecular parameters for the 1 N system. Each box of
the vv' table shows the wavelength (nm) of the v-v' transition
(upper number) and the product $B^{vv'} \cdot q_{X\rightarrow v}$
(lower number). The horizontal sum of these products is equal to the
$q_{X\rightarrow v}$ Franck-Condon factor for the molecular
excitation. Last two rows show the values of the characteristic
pressures for the quenching of the upper v level in pure nitrogen
and air (see text for more details).}

\normalsize

\scriptsize \tiny
\begin{center}
\begin{tabular}{c||c|c|c|c|c|c|c|c|c||c||c|c|}

\hline
v/\ v' & 0 & 1 & 2 & 3 & 4 & 5 & 6 & 7 & 8 & $q_{X\rightarrow v}$  & $P^{v'}_{nitr}$ (hPa) & $P^{v'}_{air}$ (hPa)\\

\hline

 & 337.0 & 357.6 & 380.4 & 405.8 & 434.3 & 466.5 & 503.2 & 545.2 &
   593.8 & &  &\\
0 \hspace{0.3cm} & 0.265 & 0.175 & 0.072 & 0.022 & 0.006 & 0.001 &
0.000 & 0.000 &
   0.000 & 0.545 & 77 & 13.1\\

\hline

 & 315.8 & 333.8 & 353.6 & 375.4 & 399.7 & 426.8 & 457.3 & 491.7 &
 530.9 & & & \\
1 \hspace{0.3cm} & 0.138 & 0.007 & 0.064 & 0.057 & 0.028 & 0.010 &
0.003 & 0.001 &
   0.000 &  0.308 & 36 & 11.2\\

\hline

 & 297.6 & 313.5 & 330.9 & 349.9 & 370.9 & 394.2 & 420.0 & 448.9 &
   481.3 & & &  \\
2 \hspace{0.3cm} & 0.016 & 0.041 & 0.003 & 0.007 & 0.016 & 0.013 & 0.006 & 0.002 & 0.001 &  0.106 & 23  & 9.1\\

\hline

 & 281.8 & 296.2 & 311.5 & 328.4 & 346.8 & 367.1 & 389.4 & 414.0 & 441.5 & & & \\
3 \hspace{0.3cm} & 0.001 & 0.009 & 0.007 & 0.003 & 0.000 & 0.003 &
0.004 & 0.002 &
   0.001 & 0.030 & 22  & 7.9 \\

\hline

 & 268.4 & 281.2 & 295.2 & 310.2 & 326.6 & 344.5 & 364.1 & 385.6 &
   409.3 & &  & \\
4 \hspace{0.3cm} & 0.000 & 0.000 & 0.003 & 0.001 & 0.001 & 0.000 &
0.000 & 0.001 &
   0.001 & 0.008  & 21  & 6.7\\
\hline

\end{tabular}
\end{center}

\small

Table 3.- {\sl Same as Table 2 for the 2 P system.}

\normalsize

Values for the involved parameters are shown in Tables 2 and 3 for
1N and 2P systems respectively. Accurate Franck-Condon factors
$q_{X\rightarrow v}$ and branching ratios $B^{vv'}$ have been taken
from \cite{gilmore}. In regard with the characteristic pressures
$P'_v$, many measurements are available in the literature.

\par

For the 1N system \cite{bunner}, \cite{nagano2} and \cite{mitchell}
provide experimental results for the v=0 level for both pure
nitrogen and dry air. A reasonable agreement between \cite{bunner}
and \cite{mitchell} is found and an average value of both results,
shown in Table 2, has been used in our calculations. Since no
results for $v>0$ seems to be available, the same $P'_0$ value has
been used for all v levels. The bands of this system are strongly
attenuated by quenching and therefore their contribution at high
pressure is very small, consequently, even a severe error in $P'_v$
($v>0$) would not have a significant effect on the calculated total
fluorescence yield.

\par

On the other hand, quenching of the 2P system by nitrogen
(auto-quenching) has been extensively studied by many authors
\cite{bunner}, \cite{nagano2}, \cite{mitchell},
\cite{becker,carr,millet,calo,nichols,urosevic,simek,Pancheshnyi,morozov,Chen,Gat,Albugues}.
Unfortunately large discrepancies between the various measurements
are found. Averaged values from \cite{simek}, \cite{Pancheshnyi} and
\cite{morozov}, in good agreement with a significant fraction of
previous measurements, have been used in this work (see Table 3).

\par

The total fluorescence yield for the bands contributing to the usual
experimental spectroscopic interval (300 - 406 nm) has been
evaluated as a function of electron energy and pressure. Note that
expression (\ref{TFY2}) had to be corrected to account for the
missed transitions (those outside the above wavelength limits). A
plot of $\varepsilon (E)$ for pure nitrogen in the interval ranging
from threshold to the GeV region at several pressures ranging from 1
hPa to atmospheric pressure is shown in Figure 8. Several
interesting features of $\varepsilon (E,P)$ can be observed in this
figure. At high pressure, expression (\ref{TFY2}) becomes

\begin{equation}
\label{TFY4} \varepsilon^{system} =
\frac{1}{kT}\frac{\sigma^{eff}_{00}(E,P) }{q_{X\rightarrow 0}B^{00}}
{\sum_{v} q_{X\rightarrow v}{P'_v}}\,,
\end{equation}

showing that for $P \gg P'_v$ the fluorescence yield is basically
independent on pressure. The only remaining dependence is the small
one from the $\alpha$ parameter and (even smaller) that from the
density correction of $\sigma_{ion}$ at high energy. That is, the
increase of fluorescence with pressure is nearly canceled out by the
corresponding increased quenching.
Note that the higher is $P'_v$, the weaker is the pressure
dependence of fluorescence yield.

\begin{figure}[htb]
\centering \epsfig{file=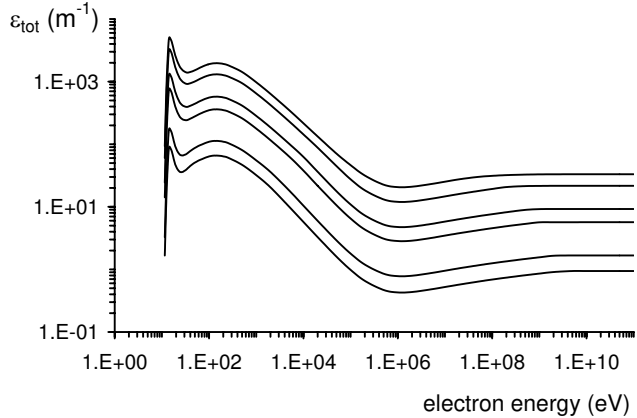,width=0.8\linewidth} \vskip
-1.5cm \caption{{\sl Total fluorescence yield (300 - 406 nm) versus
energy for pure nitrogen at pressures (bottom - up) 1, 2, 10, 20,
100 and 1013 hPa for a characteristic size of the interaction region
of $R$ = 2.5 cm.}}
\end{figure}

The $\sigma^{eff}_{00}$ parameter can be easily evaluated. For
energies about 100 eV and over, the optical cross sections of the 1N
system are proportional to the total ionization cross section,
$\sigma_{00}^{1N} (E) = \chi_{00} \sigma_{ion}(E)$ and therefore

\begin{equation}
\label{TFY5} \sigma_{00}^{1N,eff} (E, P) = (\chi_{00} +
\alpha_{00}^{1N})\sigma_{ion}\,.
\end{equation}

In regard with the 2P system the contribution of direct excitation
at high energy and high pressure ($E \gtrsim 10^3 eV $ and $P\times
R \gtrsim$ 3hPa $\times$ cm) is negligible because under these
conditions $\sigma_{vv'}\ll\alpha_{vv'}\sigma_{ion}$ and thus

\begin{equation}
\label{TFY6} \sigma_{00}^{2P,eff} (E, P) \approx
\alpha_{00}^{2P}\sigma_{ion}\,.
\end{equation}

The $\alpha (E,P)$ parameter reaches saturation (i.e. it becomes
independent of $E$ for increasing energies) at an energy which grows
with pressure (\ref{alpha_00}). The data in Table 1 shows that, at
atmospheric pressure, this saturation energy results of about
1.3$\times$10$^5$ eV for both systems. As a consequence for higher
energies $\varepsilon(E)$ is proportional to $\sigma_{ion} (E)$.

\begin{figure}[htb]
\centering \epsfig{file=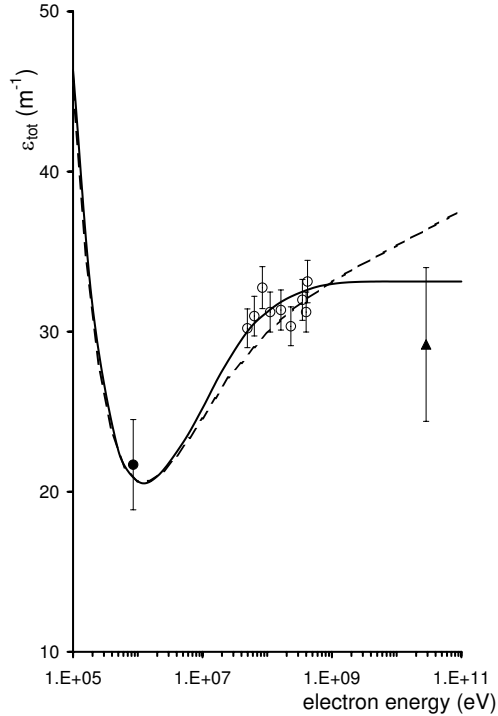,width=0.6\linewidth} \vskip
-0.5cm \caption{{\sl Total fluorescence yield (300 - 406 nm) at high
energy for pure nitrogen. Predictions of this work (continuous line)
are compared with absolute measurements of Nagano et al.
\cite{nagano2} ($\bullet$) and the FLASH collaboration \cite{flash}
($\blacktriangle$). Relative measurements of AIRFLY ($\circ$)
normalized to our predictions at 10$^8$ eV are also shown. Dashed
line represents the Bethe-Bloch function normalized at 1 MeV.}}
\end{figure}

The results of our calculations have been compared with available
experimental data. Figure 9 shows the total fluorescence yield (300
- 406 nm) at high energy and atmospheric pressure for pure nitrogen
together with the absolute measurements of Nagano et al.
\cite{nagano2} at 0.850 MeV and those of the FLASH collaboration
\cite{flash} at 28.5 GeV. In addition, relative measurements of
AIRFLY \cite{air-fly} in the interval 50 - 500 MeV have been
represented. As shown in the figure, experimental data for pure
nitrogen are in good agreement (within the experimental errors) with
our predictions.

\par

Unfortunately, quenching measurements in air for the 2P system are
scarce, showing also large discrepancies. Bunner \cite{bunner} gives
$P'_v$ values of 20 and 8.7 hPa for $v$ = 0 and 1 respectively.
Nagano et al \cite{nagano2} measure $P'$ values from the fitting of
experimental $\varepsilon(P)$ results. Average values of about 18,
25 and 8 hPa are found for $v$ = 0, 1 and 2 respectively. As already
mentioned these $P'_v$ values could be overestimated by the
contribution of secondary electrons. Finally, Pancheshnyi et al.
\cite{Pancheshnyi} have published measurements of the Nitrogen 2P
quenching by O$_2$ molecules in good agreement with the results of
\cite{Albugues} and \cite{millet}. These measurements give the air
$P'_v$ values shown in Table 3 which have been used in our
calculations.

\begin{figure}[htb]
\centering \epsfig{file=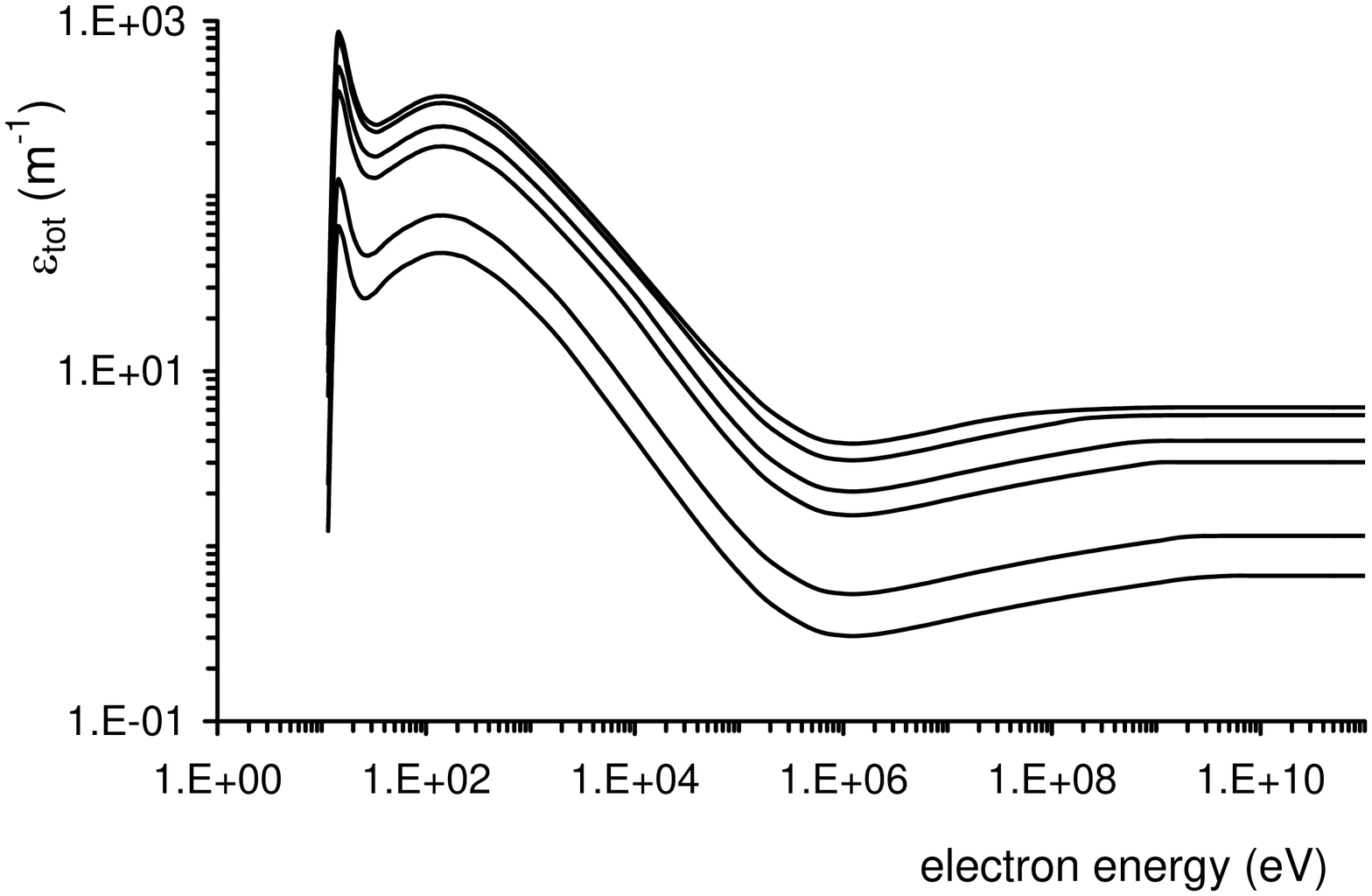,width=0.9\linewidth} \vskip
-1.5cm \caption{{\sl Total fluorescence yield (300 - 406 nm) for dry
air against electron energy at pressures (bottom-up) 1, 2, 10, 20,
100, 1013 hPa for a characteristic size of the interaction region of
$R$ = 2.5 cm.}}
\end{figure}

Figure 10 shows plots of $\varepsilon (E)$ for dry air in a wide
energy interval for several pressures ranging from 1 hPa to
atmospheric pressure. In Figure 11 these results are compared with
available measurements at high energy. Our calculations are in good
agreement with that of Nagano et al. \cite{nagano2} at 0.850 MeV but
they predict a fluorescence yield about 20\% larger than the
measurements of Kakimoto et al. \cite{kakimoto} (at several energies
in the range 1 MeV - 1GeV) and those of the FLASH collaboration
\cite{flash} at 28.5 GeV.

\begin{figure}[htb]
\centering \epsfig{file=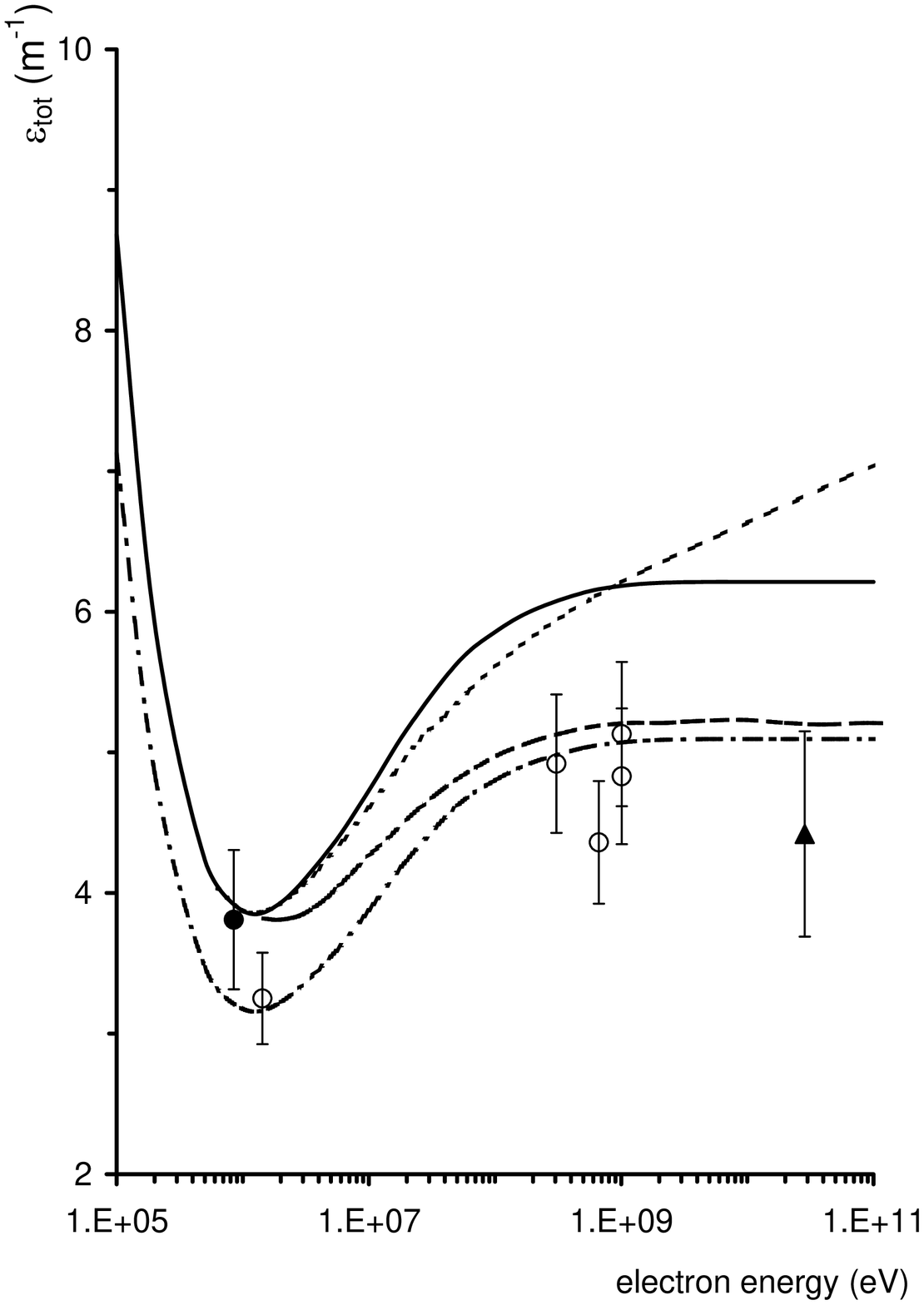,width=0.5\linewidth} \vskip
-1.0cm \caption{{\sl Total fluorescence yield (300 - 406 nm) at high
energy for dry air. Continuous line represents the prediction of
this work using the parameters of Tables 2 and 3. The comparison
with experimental results of Nagano et al. \cite{nagano2}
($\bullet$), Kakimoto et al. \cite{kakimoto} ($\circ$) and the FLASH
collaboration \cite{flash} ($\blacktriangle$) suggests that $P'_v$
values could be overestimated (see text for more details). Results
of this work reduced by 18\% (dash-dotted line) are in good
agreement with all measurements. Dotted line represents the stopping
power and the dashed line is the deposited energy from \cite{flash}
both normalized to $\varepsilon$ at 1 MeV.}}
\end{figure}

The calculations presented in this work relies in several molecular
parameters and in a model for evaluating the contribution of
secondary electrons. Branching ratios, Franck-Condon factors and the
validity of the Franck-Condon principle are reliable and thus their
contribution to the uncertainty in $\varepsilon$ is expected to be
very small. On the other hand, at high pressure, fluorescence yields
are proportional to $P'_v$ values for which large discrepancies are
found in the available measurements. The uncertainty in these
parameters are very likely one of the main error sources in our
calculations. On the other hand, our model \cite{blanco,blanco2}
uses several molecular parameters which can also contribute to a
systematic error in the evaluation of the fluorescence light from
secondary electrons, in particular, the absolute value of the total
excitation cross section of N$_2$. This uncertainty leads to a
possible (small) constant factor error. In addition, the energy
spectrum of secondary electrons may affect the energy dependence of
the fluorescence yield. In this work the spectrum used in
\cite{blanco} has been assumed. On the contrary, the geometry of the
interaction volume (different for the various experiments) should
not contribute significantly to our errors since $\alpha_{vv'}$
grows very slowly with $R$ ($\approx \ln R$). In regard with the
energy dependence, our calculations rely in the validity of the well
established Born-Bethe approximation at high energy.

\par

In summary, at high pressure the predictions of this work could be
affected by a small error (constant factor) mainly due to the
uncertainty in the quenching efficiency together with a possible
error in the energy dependence due to inaccuracies in the spectrum
of secondary electrons. In Figure 11 our fluorescence yield reduced
by a constant factor of 18\% has been plotted showing a reasonable
agreement with all experimental results. Note that the ratio of
fluorescence yields $\varepsilon^{air}/\varepsilon^{nitrogen}$ in
experiments measuring both gases (air and nitrogen) at high pressure
are expected to be equal to the ratio $P'_{air}/P'_{nitrogen}$ since
experimental errors (calibrations, geometrical factors, etc.) are
the same for both gases. Our predictions for nitrogen agree with the
absolute measurements of \cite{nagano2} and \cite{flash} while a
fluorescence efficiency about 20\% larger than \cite{kakimoto} and
\cite{flash} is predicted for air. This lead us to suspect that the
ratio $P'_{air}/P'_{nitrogen}$ assumed in our calculations has been
overestimated by about a 20\%.

\par

In Figures 9 and 11 a plot of the stopping power law for electrons
(Bethe-Bloch) has been represented showing a noticeable deviation
with respect to $\varepsilon (E)$ as predicted, since at large $E$
values a significant fraction of the electron energy loss is not
deposited inside the fluorescence cell. At low energy the
fluorescence yield also deviates from the stopping cross section as
shown in Figure 12. As already mentioned most of the fluorescence
light is generated by low energy secondary electrons for which
deposited energy is proportional to the stopping cross section.
Since at low $E$, $\varepsilon/\sigma_{st}$ is energy dependent, the
$\varepsilon$ to deposited energy ratio is expected to depend on $E$
at high energy also. In Figure 11 the deposited energy versus $E$
reported by \cite{flash} for the FLASH experiment has been also
represented (normalized to our fluorescence yield at 1 MeV). Our
calculations predict that at large energies a growing fraction of
deposited energy is converted in fluorescence radiation. According
to this figure, the ratio $\varepsilon$ to deposited energy grows
about 19\% in the interval 1 MeV - 1 GeV.

\begin{figure}[htb]
\centering \epsfig{file=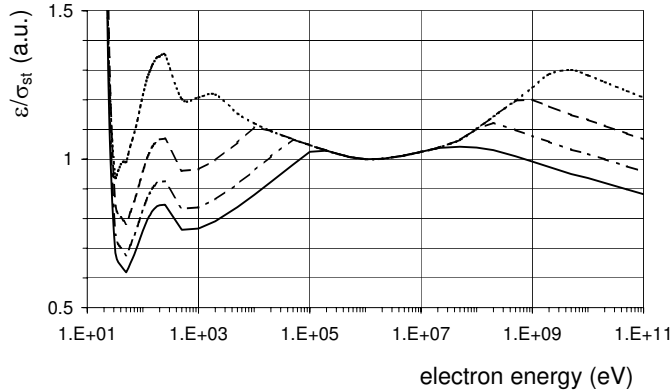,width=0.8\linewidth} \vskip
-1.0cm \caption{{\sl Ratio of fluorescence yield over stopping cross
section (Bethe-Bloch) for air (arbitrary units) at 1 hPa (dotted),
20 hPa (dashed), 200 hPa (dot-dashed) and 1013 hPa (continuous). The
ratio has been normalized to 1 at 1MeV.}}
\end{figure}

\section{Conclusions}
\label{conclu}
Theoretical predictions on the fluorescence generation of nitrogen
molecules excited by electrons have been presented. A general
procedure for the calculation of fluorescence yields is shown which
can be applied in a very wide energy interval ranging from threshold
up to the GeV region and for any environmental condition (pressure,
temperature, contaminants, etc.) as far as the involved molecular
parameters are available. As an example, plots of fluorescence
yields (as expected from typical laboratory experiments) versus
energy in a wide pressure range are shown.

\par

The contribution of direct excitation of 1N system is calculated
using low energy measurements of the optical cross section with
extrapolation to high energies as given by the Born-Bethe law.
Direct excitation of 2P system is negligible. Particular attention
has been paid to the the contribution of secondary electrons ejected
in ionization processes. A simple model is used to evaluate the
fluorescence from secondary electrons which turns out to be
non-negligible for the excitation of the 1N system and dominant for
the generation of the 2P fluorescence. At high energy and high
pressure fluorescence yield results proportional to the total
ionization cross section which also follows a Born-Bethe energy law.
Quenching of fluorescence by N$_2$ and O$_2$ molecules is taken into
account using available data in the literature. At high pressure,
fluorescence yield is proportional to the $P'_v$ parameters.
Unfortunately values of these parameters reported in the literature
show large disagreements.

\par

Comparison of our predictions with available measurements at high
energy ($E>$ 1MeV) and high pressure shows very good agreement for
pure nitrogen while some discrepancies of about 20\% are found for
dry air. These discrepancies are very likely due to uncertainties in
the quenching cross sections.

\par

The relationship between fluorescence yield, stopping power and
energy deposited by the electrons in the medium has been discussed.
In a typical laboratory experiment, at very high energy, many
secondary electrons reach the wall of the cell before emitting all
the fluorescence light and therefore stopping power grows with
energy faster than fluorescence yield. On the other hand, most of
the fluorescence light is generated by low energy secondary
electrons for which the optical cross section is not proportional to
the stopping power and therefore, in principle, proportionality
between fluorescence yield and deposited energy is not assured. A
comparison of our fluorescence yield results with deposited energy
in the FLASH experiment shows a deviation from proportionality of
about 20\% in the interval 1 MeV - 10 GeV. Note that the uncertainty
of this result depends on possible inaccuracies in the energy
spectrum of secondary electrons assumed in our model \cite{blanco}.

\par

More details on the model used in this work for the evaluation of
the fluorescence contribution of secondary electrons will be
published soon. On the other hand a detailed study on the
relationship between deposited energy and fluorescence yield
following our model is underway.

\par\vspace{1.5cm}
\noindent
\large
{\bf Acknowledgments}
\normalsize

This work has been supported by the Spanish Ministry of Science and
Education MEC (ref.: FPA03-08733-C02-01). A.Castellanos acknowledges
an undergraduate grant from the program ``Becas Colaboraci\'on'' of
MEC

\end{document}